\documentclass[bib]{statapress}

\usepackage[crop,newcenter,frame]{pagedims}
\usepackage{sj,epsfig,stata,shadow}

\sjsetissue{$vv$}{$ii$}{$mm$}{$yyyy$}

\newcommand{\FWER}{\mathrm{FWER}}
\newcommand{\BetaDist}{\mathrm{Beta}}
\newcommand{\UnifDist}{\mathrm{Unif}}

\newcommand{\iid}{\stackrel{\mathrm{iid}}{\sim}}

\begin{document}

% SJ EDITORS will change this [right?]
\inserttype[st0001]{article}

\author{D.\ M.\ Kaplan}{David M.\ Kaplan\\Department of Economics, University of Missouri\\Columbia, MO, USA\\kaplandm@missouri.edu}

\title[distcomp]{distcomp: Comparing distributions}

\maketitle

\begin{abstract}
The \stcmd{distcomp} command is introduced and illustrated.
The command assesses whether or not two distributions differ at each possible value while controlling the probability of any false positive, even in finite samples.
Syntax and the underlying methodology \citep[from][]{GoldmanKaplan2018c} are discussed.
Multiple examples illustrate the \stcmd{distcomp} command, including revisiting the experimental data of \citet{GneezyList2006} and the regression discontinuity design of \citet*{CattaneoEtAl2015}.

\end{abstract}

\section{Introduction}
\label{sec:intro}

The new \stcmd{distcomp} command implements a new statistical procedure for comparing distributions, introduced in \citet{GoldmanKaplan2018c}.
The usage is similar to a two-sample $t$-test or two-sample Kolmogorov--Smirnov test, i.e., \stcmd{ttest} or \stcmd{ksmirnov} (respectively) with the \texttt{by} option (see \rref{ttest} or \rref{ksmirnov}).
However, instead of only comparing the distributions' means (like \stcmd{ttest}) or only testing a single hypothesis of distributional equality (like \stcmd{ksmirnov}), \stcmd{distcomp} assesses equality of the distribution functions point by point.
Thus, instead of a single rejection or non-rejection, \stcmd{distcomp} displays ranges of values in which the distributions' difference is statistically significant.
(Goodness-of-fit test results like from \stcmd{ksmirnov} can also be shown.)

The new procedure controls false positives with a property called strong control of the familywise error rate (described later).
As a special case, if the distributions are truly identical, then no ranges will be deemed statistically significant $95\%$ of the time if a $5\%$ level is used.
This familywise error rate is controlled in finite samples (not just asymptotically).

Even for goodness-of-fit testing, \stcmd{distcomp} may be preferred to \stcmd{ksmirnov} since the new method's sensitivity to deviations is more evenly spread across the distribution.
This new goodness-of-fit test was proposed by \citet{GoldmanKaplan2018c}, refining an idea from \citet{BujaRolke2006}.
Specifically, the Kolmogorov--Smirnov test has long been known to lack sensitivity to deviations in the tails of a distribution \citep[e.g.,][p.\ 117]{Eicker1979}.
For example, if one sample has observed values $0.02,0.04,\ldots,0.98$, like a standard uniform distribution, the second sample may have even six out of $21$ values exceeding one million without a two-sided Kolmogorov--Smirnov test rejecting at a $10\%$ level.
% R code:  ks.test(c(1:14/21,10^6+1:6), 1:49/50)
In contrast, \stcmd{distcomp} rejects equality at even a $1\%$ level.
The following Stata code shows such a result.

% \onnextpage
%[auto]
\begin{stlog}
. set obs 69
number of observations (_N) was 0, now 69
{\smallskip}
. gen grp = (_n>49)
{\smallskip}
. gen y = (_n<=49)*(_n/50) + (_n>49)*(_n-49)/21
{\smallskip}
. replace y = 1000000+_n if _n>63
(6 real changes made)
{\smallskip}
. ksmirnov y , by(grp) exact

Two-sample Kolmogorov-Smirnov test for equality of distribution functions
{\smallskip}
 Smaller group       D       P-value      Exact
 ----------------------------------------------
 0:                  0.3000    0.078
 1:                 -0.0378    0.960
 Combined K-S:       0.3000    0.155      0.121
{\smallskip}
. distcomp y , by(grp) alpha(0.01) noplot p
Comparing distribution of y when grp=0 vs. grp=1
 
Global test of equality of two CDFs:
    Simulated p-value = .0055889
    At a 10\% level: reject
    At a  5\% level: reject
    At a  1\% level: reject
\end{stlog}

Section \ref{sec:gentle} discusses the methodology at a relatively intuitive level.
Section \ref{sec:distcomp} describes syntax and usage of \stcmd{distcomp}.
Section \ref{sec:ex} provides empirical examples that can be replicated with the provided do-file.
Section \ref{sec:methodology} shows some of the theoretical foundations before concluding.
Readers interested in related methods (one-sided, one-sample, or uniform confidence bands) are referred to \citet{GoldmanKaplan2018c} and the corresponding R code.
Abbreviations are used for cumulative distribution function (CDF), goodness-of-fit (GOF), Kolmogorov--Smirnov (KS), and familywise error rate (FWER).

\section{A gentle introduction to methodology}
\label{sec:gentle}

This section discusses methodology at a relatively non-technical level (compared to Section \ref{sec:methodology}).
Interest is in the distribution of some outcome variable (like wage) for two groups (like union members and non-members).
More specifically, the question is whether (and where) the two cumulative distribution functions (CDFs) are different.
Let $F(\cdot)$ be the first group's CDF (like the CDF of wage for union members); $G(\cdot)$ is the second group's CDF.
Estimated CDFs $\sthat{F}(\cdot)$ and $\sthat{G}(\cdot)$ are computed from iid samples.
These are the stair-step functions like in the graphs in Section \ref{sec:ex}.

Consider the Kolmogorov--Smirnov (KS) goodness-of-fit (GOF) test.
The null hypothesis is
\begin{equation}\label{eqn:H0-GOF}
\textrm{(GOF)}
\quad
H_0 \colon F(r)=G(r)\textrm{ for all }r ,
\end{equation}
identical CDFs.
If $H_0$ is true, then $\sthat{F}(\cdot)$ and $\sthat{G}(\cdot)$ should be ``close'' to each other; if not, the test rejects.

KS defines ``close'' with vertical distance.
At point $r$, this distance is
\begin{equation}\label{eqn:Dhat}
\sthat{D}(r) \equiv \vert \sthat{F}(r)-\sthat{G}(r) \vert .
\end{equation}
To compare the entire functions, KS looks across all $r$ to find the biggest gap, $\max_r\sthat{D}(r)$.
Happily, under $H_0$, the sampling distribution of the ``biggest gap'' does not depend on the true distribution, so it can be simulated.
Thus, finite-sample $p$-values can be simulated, without any asymptotic approximation, and the KS test can control the type I error rate at level $\alpha$.

Instead of (\ref{eqn:H0-GOF}), null hypotheses can be defined to show \emph{where} two distributions differ.
For each possible value $r$ of the outcome variable, define
\begin{equation}\label{eqn:H0r}
H_{0r} \colon F(r)=G(r) .
\end{equation}
The GOF null hypothesis could be rewritten as
\begin{equation}\label{eqn:H0-GOF2}
\textrm{(GOF again)}
\quad
H_0 \colon \textrm{all }H_{0r}\textrm{ are true} .
\end{equation}
Whereas the GOF test only distinguishes whether all $H_{0r}$ are true or at least one is false, now we care about specifically which $H_{0r}$ are true and which are false.
This is summarized by \stcmd{distcomp} by the ranges of $r$ for which $H_{0r}$ is rejected.

It does not work to run a level $\alpha$ $t$-test on all $H_{0r}$.
Imagine just two values $r=1,2$, and $\alpha=0.1=10\%$.
Let both $H_{0r}$ be true.
Then, each test's probability of not rejecting is $0.9$.
Assuming independence for simplicity, the probability that neither rejects is $(0.9)(0.9)=0.81$, so the probability that one or both rejects is $1-0.81=0.19$.
That is, the probability of at least one type I error (false positive) is $19\%$, much larger than the desired $\alpha=10\%$.
This is called the ``multiple testing problem.''

The following is one way to define type I error control for multiple testing.
A ``familywise error'' is made if at least one true $H_{0r}$ is rejected.
For example, if $H_{0r}$ is true only for $r\le0$, then a familywise error occurs if any $H_{0r}$ with $r\le0$ is rejected.
The probability of such an error is the familywise error rate (FWER):
\begin{equation}\label{eqn:FWER}
\FWER \equiv \Pr(\textrm{reject any true }H_{0r}) .
\end{equation}
``Weak control of FWER at level $\alpha$'' guarantees $\FWER\le\alpha$ when all $H_{0r}$ are true.
The \stcmd{distcomp} methodology achieves ``strong control'' of FWER, meaning $\FWER\le\alpha$ regardless of which $H_{0r}$ are true.
Put differently: with strong control of FWER at a $10\%$ level, there will be zero false positives $90\%$ of the time.

One way to test (\ref{eqn:H0r}) and achieve strong control of FWER is to extend the KS approach.
Specifically, imagine that $c$ is the simulated critical value given $n$ and $\alpha$.
It seems natural to reject $H_{0r}$ when $\sthat{D}(r)>c$.

The weak control of FWER comes from the KS GOF test's properties.
When the GOF $H_0$ is true (so all $H_{0r}$ are true), the probability of GOF rejection is $\alpha$: $\Pr(\max_r\sthat{D}(r)>c)=\alpha$.
Since $\max_r\sthat{D}(r)>c$ is equivalent to ``at least one $H_{0r}$ is rejected,'' the probability of rejecting at least one $H_{0r}$ (i.e., FWER) is also $\alpha$.

The KS approach also has strong control of FWER.
The intuition is that when all $H_{0r}$ are true (as with weak control of FWER), the opportunity for a familywise error is greatest, so FWER is highest.
If fewer $H_{0r}$ are true, then there is less opportunity for a familywise error.
In the extreme, when no $H_{0r}$ is true, it is impossible to make a familywise error.
This intuition is proved correct for a certain class of methods including the KS approach and \stcmd{distcomp}, although it can fail in other cases.

The KS approach controls FWER, but it distributes power unevenly.
This means it detects certain types of differences between $F(\cdot)$ and $G(\cdot)$ very well, but detects others very poorly.
Unless we know what type of difference to expect, it seems prudent to desire power against all differences.
This is achieved by \stcmd{distcomp}, without sacrificing the finite-sample strong control of FWER.

One shortcoming of KS is its implicit symmetry.
The definition of $\sthat{D}(r)$ has an absolute value; positive and negative values are treated the same.
This makes sense with normal distributions, but normality is only an asymptotic approximation.
The finite-sample distribution of $\sthat{F}(r)$ is actually binomial, which is especially skewed (not symmetric) when $r$ is in the tails.

KS also ignores scale differences.
The variance of $\sthat{F}(r)$ is highest when $r$ is the median, and nears zero when $r$ is in the tails.
The variance of $\sthat{D}(r)$ similarly varies with $r$.
Even if $\sthat{D}(r)$ were normal, it would be much more likely have a large value with median $r$ than with $r$ in the tails, where $H_{0r}$ is very unlikely to be rejected.

By using the finite-sample sampling distributions of $\sthat{F}(r)$ and $\sthat{G}(r)$, \stcmd{distcomp} accounts for both skewness and scale differences across $r$, unlike KS.

\section{The distcomp command}
\label{sec:distcomp}

The \stcmd{distcomp} command compares two distributions.
One variable in the data (\varname\ below) is the variable for comparison, like price or income.
Another variable (\textit{groupvar} below) takes only two distinct values, defining two groups, like an indicator/dummy for male whose value is $0$ or $1$, or a state abbreviation whose value is NY or CA.

The validity of two assumptions should be considered in practice.
First, sampling is assumed independent and identically distributed (iid) from the two respective group population distributions, and it is assumed the groups are sampled independently.
Second, the variable of interest (\varname) is assumed to have a continuous distribution, but some amount of discreteness is ok.
In particular, if there are duplicate values within each sample, but no ``ties'' (same value observed in both samples), then the properties remain the same.
However, if there are many ties, then the theoretical results do not apply directly and the properties may change substantially.
In the absence of theoretical results allowing ties, simulations suggest the method may become conservative, controlling the FWER at a level even lower than the level specified.
One such simulation is included in the accompanying \texttt{distcomp\_examples.do} file, in which the nominal FWER is $10\%$ but simulated FWER is near $0\%$.

The first results displayed by \stcmd{distcomp} are for the global (goodness-of-fit, GOF) test.
This is the same type of test as the (two-sample) Kolmogorov--Smirnov test.
That is, the null hypothesis is that the two cumulative distribution functions (CDFs) are identical.
This could be false even if the two distributions' means are identical (and \stcmd{ttest} does not reject), e.g., with normal distributions with the same mean but different standard deviation.
The global test results are always reported for levels $1\%$, $5\%$, and $10\%$.
Optionally, a $p$-value is also reported; it must be simulated and can substantially increase computation time with large sample sizes.
(To avoid misinterpretation, when the simulated value is zero, $0.0001$ is reported since there are $10^4$ simulation replications.)
The test is generally more powerful than Kolmogorov--Smirnov.
The GOF methodology was proposed by \citet{GoldmanKaplan2018c}, refining an idea from \citet{BujaRolke2006}.

The second results displayed are for a multiple testing procedure.
They show ranges of values for which the difference between CDFs is statistically significant, accounting for the multiple testing nature of the procedure (i.e., many different points are tested simultaneously).
Instead of a single, GOF null hypothesis, there is a set of many null hypotheses.
Within the set, each individual hypothesis specifies equality of the two CDFs at a different point.
That is, if $F(\cdot)$ and $G(\cdot)$ are the two CDFs, then each individual null hypothesis is $H_{0x} \colon F(x)=G(x)$, and the set of such hypotheses for all possible values of $x$ is considered.
The multiple testing procedure rejects equality at certain values of $x$ while controlling the probability of \emph{any} type I error (false positive).
The probability of any false positive is known as the familywise error rate (FWER).
The \stcmd{distcomp} procedure controls the finite-sample (not just asymptotic) FWER at the desired level specified by the user.
The output shows the ranges of $x$ where $H_{0x} \colon F(x)=G(x)$ is rejected.
This methodology is from \citet{GoldmanKaplan2018c}.

By default, a plot is generated with the empirical CDFs of the two groups, along with the rejected ranges (if any).

The restriction of FWER levels to $1\%$, $5\%$, and $10\%$ allows nearly instantaneous computation (when a $p$-value is not requested).
The reason is that a table of precise ``critical values'' for these specific levels has been simulated ahead of time.
In small samples, as with the Kolmogorov--Smirnov test, often it is impossible to attain exactly $1\%$, $5\%$, or $10\%$.
For example, for certain sample sizes, it may only be possible to have FWER of $9.6\%$ or $10.6\%$, but nothing in between.
To make this transparent, the exact finite-sample FWER level has also been pre-simulated and is returned by \stcmd{distcomp}.
In some cases, an analytic formula (based on simulations) is used, in which case the pre-simulated exact FWER is not available.
Except with very small samples, the practical difference between specified and actual FWER level is usually negligible.

Syntax, options, and stored results are now shown; prefix \stcmd{by} is allowed.

\begin{stsyntax}
distcomp
  \varname\
  \optif\
  \optin\
  , by(\textit{groupvar})
  \optional{\underbar{a}lpha(\num) \dunderbar{p}value noplot}
\end{stsyntax}

\hangpara
\texttt{by(}{\textit{groupvar}\texttt{)} is required.
It specifies a binary variable that identifies the two groups whose distributions are compared.
(The variable does not need to have values $0$ and $1$ specifically; any two values are fine, like $1$ and $5$, or ``cat'' and ``dog.'')

\hangpara
\texttt{alpha(\num)} specifies the familywise error rate (FWER) level, as a decimal.
The default is $0.10$, i.e., $10\%$ probability of any false positive.
Other accepted values are $0.05$ and $0.01$ (meaning $5\%$ and $1\%$).

\hangpara
\texttt{pvalue} 
requests a global $p$-value, computed by simulation.
The computation time depends on the sample size.
For prohibitively large samples, the global test's rejections at levels $1\%$, $5\%$, and $10\%$ (which do not require simulation) can be used to roughly approximate the $p$-value.
For example, if only the $10\%$ test rejects, then $0.05<p<0.10$.

\hangpara
\texttt{noplot}
suppresses the plot.

% \stcmd{distcomp} stores the following in \texttt{r()}:

\begin{stresults}
\stresultsgroup{Scalars} \\
\stcmd{r(rej\_gof10)} & GOF rejection, $10\%$ level &
\stcmd{r(p\_gof)} & GOF $p$-value \\
\stcmd{r(rej\_gof05)} & GOF rejection, $\phantom{1}5\%$ level &
\stcmd{r(alpha\_sim)} & simulated FWER level \\
\stcmd{r(rej\_gof01)} & GOF rejection, $\phantom{1}1\%$ level &
\stcmd{r(alpha)} & specified FWER level \\
\\\stresultsgroup{Matrices} \\
\stcmd{r(N)} & numbers of observations &
\stcmd{r(rej\_ranges)} & ranges with CDF equality rejected
\end{stresults}

Certain stored results may be missing in some cases.
When \texttt{r(alpha\_sim)} is not applicable, it is set to missing.
The GOF $p$-value is only computed if requested.
When no ranges are rejected, the \texttt{r(rej\_ranges)} is a 1-by-2 matrix with both entries missing.

Also, \stcmd{r(N)} shows three numbers; in order: the overall, first group, and second group number of observations used for analysis.

\section{Examples}
\label{sec:ex}

The examples in this section can all be replicated with the file \texttt{distcomp\_examples.do}.
Some code is omitted here to conserve space.

\subsection{Simple example with built-in dataset}
\label{sec:ex-NLSW}

The following example compares the hourly wage distributions of union and non-union workers in the NLSW 1988 extract shipped with Stata, with a $10\%$ statistical significance level.

% [auto]
% \onnextpage
\begin{stlog}
. sysuse nlsw88 , clear
(NLSW, 1988 extract)
{\smallskip}
. set scheme sj
{\smallskip}
. distcomp wage , by(union) alpha(0.10) p
Comparing distribution of wage when union=0 vs. union=1
 
Global test of equality of two CDFs:
    Simulated p-value < .0001
    At a 10\% level: reject
    At a  5\% level: reject
    At a  1\% level: reject
 
With strong control of FWER at a 10\% level,
CDF equality is rejected at all points in the following
ranges of wage:
{\smallskip}
     from         to 
 2.383252   2.520128 
 2.568437   11.51368 
 11.52979    11.5781 
\end{stlog}

Three main results are displayed.
The first result says that the global/GOF null hypothesis that the two wage distributions are identical is rejected at a $1\%$ (and thus $5\%$ and $10\%$) level.
The statistical significance is even greater since the $p$-value is even lower than $0.01$.
There is very strong evidence that union and non-union wage distributions are not identical.
The second result shows the range of wage values at which CDF equality is rejected while controlling the FWER at $10\%$.
In this example, the range covers most of the wage distribution, from around the $2$nd percentile ($\$2.38$/hr) to almost the $86$th percentile ($\$11.61$/hr).
This suggests a restricted first-order stochastic dominance relationship, as defined in Condition I of \citet[p.\ 751]{Atkinson1987}.
The empirical CDF graph (Figure \ref{fig:wage}) illustrates this result, too.

\begin{figure}[htbp]
\centering
\epsfig{file=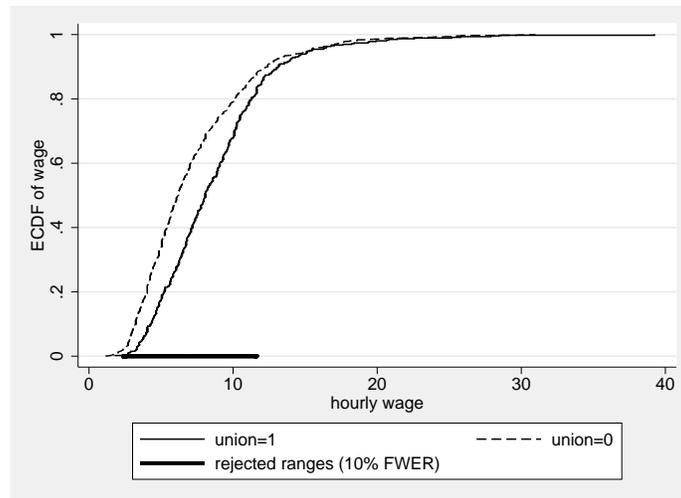}
\caption{\label{fig:wage}Empirical CDFs from NLSW.}
\end{figure}

Figure \ref{fig:wage} shows the default plot generated by \stcmd{distcomp}.
It shows the empirical CDFs for union wage and non-union wage.
These are step functions, but each step is very small since the sample size is moderate ($461$ union, $1417$ non-union).
The thick horizontal line near the bottom shows the ranges where CDF equality was rejected ($10\%$ FWER level).
It is clear that union wages tend to be higher in the sample (hence the union CDF lies below the non-union CDF).
However, the empirical CDFs alone cannot show statistical significance.
The line at the bottom shows that this difference is indeed statistically significant at a $10\%$ FWER level across most of the distribution, though not the upper tail.

Union and non-union wages can be compared for certain subgroups, with an \stcmd{if} condition or \stcmd{by} prefix.
For example, the above analysis can be repeated for each different \stcmd{race} group with \stcmd{bysort race : distcomp wage , by(union) p} or for only married individuals with \stcmd{distcomp wage if married==1 , by(union) p}.

\subsection{Example with simulated data}
\label{sec:ex-sim}

The following example uses simulated data.
The code (including the random seed) to replicate the simulated data is in \texttt{distcomp\_examples.do} but omitted here.
The control group sample has $50$ observations drawn independently from a standard normal distribution.
The first treatment group also has $50$ observations drawn independently from a standard normal distribution, i.e., the treatment and control population distributions are identical (but sample values differ).
The second treatment group is the same for below-median individuals, but the treatment increases the outcome by two units for individuals with above-median values.
The third treatment has no effect on the mean, but it affects the spread: values are drawn from a mean-zero normal distribution with standard deviation of three.

Running \stcmd{distcomp} produces the following results.

% [auto]
\begin{stlog}
. distcomp y1 , by(treated) p // no effect
Comparing distribution of y1 when treated=0 vs. treated=1
 
Global test of equality of two CDFs:
    Simulated p-value = .395
    At a 10\% level: do not reject
    At a  5\% level: do not reject
    At a  1\% level: do not reject
{\smallskip}
. distcomp y2 , by(treated) p // effect only above median (zero)
Comparing distribution of y2 when treated=0 vs. treated=1
 
Global test of equality of two CDFs:
    Simulated p-value < .0001
    At a 10\% level: reject
    At a  5\% level: reject
    At a  1\% level: reject
 
With strong control of FWER at a 10\% level,
CDF equality is rejected at all points in the following
ranges of y2:
{\smallskip}
     from         to 
 .9370626    3.08856 
{\smallskip}
. distcomp y3 , by(treated) p // bigger in tails, 0 at median
Comparing distribution of y3 when treated=0 vs. treated=1
 
Global test of equality of two CDFs:
    Simulated p-value = .0051
    At a 10\% level: reject
    At a  5\% level: reject
    At a  1\% level: reject
 
With strong control of FWER at a 10\% level,
CDF equality is rejected at all points in the following
ranges of y3:
{\smallskip}
     from         to 
-3.862472  -1.974054 
-1.279222  -1.172997 
 1.558096   3.265678 
\end{stlog}

Above, for the \stcmd{y1} case where the control and treatment distributions are equal, nothing is statistically significant at conventional levels.
The \stcmd{distcomp} GOF test does not reject equality even at a $10\%$ level (the GOF $p$-value is $0.395$), so no ranges are rejected at $10\%$ FWER level.
The empirical CDFs differ, but \stcmd{distcomp} says these differences are not statistically significant at a $10\%$ level.

For the \stcmd{y2} case where the treatment has an effect, but only above the median (which is zero), the \stcmd{distcomp} results reflect this.
First, equality of the distributions is rejected at a level well below $1\%$ ($p<0.0001$).
Then, more specifically, \stcmd{distcomp} says equality is rejected over the range $[0.937, 3.089]$.
The distributions indeed differ over this range.
They actually differ over the larger range from zero to infinity, but there is not enough data to be certain that differences closer to zero are statistically significant, and similarly for differences far in the upper tail (above $3.089$).

For the \stcmd{y3} case where the treatment affects the standard deviation, the true CDFs differ everywhere except at zero, and again \stcmd{distcomp} reflects this.
In addition to rejecting global equality at a $1\%$ level, \stcmd{distcomp} identifies three specific ranges of values (that exclude zero) where the distributions differ.
Similar to the second case, it is most difficult to infer a difference near zero (where the CDFs are actually equal) and far in the tails (where there are few/no observations).
Given the same FWER level, more data would be required to enlarge the ranges where we are statistically confident in a CDF difference.

\subsection{Example with experimental data}
\label{sec:ex-exp}

The following example uses data from \citet{GneezyList2006} to test for distributional treatment effects.
A longer version appears in \citet[\S8.1]{GoldmanKaplan2018c}.
In brief, \citet{GneezyList2006} paid control group individuals an advertised hourly wage and treatment group individuals an unexpectedly larger ``gift'' wage upon arrival.
The ``gift exchange'' question from behavioral economics is whether the higher wage induces higher effort in return.
The experiment is run separately for library data entry and door-to-door fundraising tasks.
The sample sizes are very small: $10$ and $9$ for control and treatment (respectively) for the library task, and $10$ and $13$ for fundraising.
With small samples, the finite-sample FWER control of \stcmd{distcomp} is especially desirable.
Complementing the original results of \citet{GneezyList2006}, \citet{GoldmanKaplan2018c} examine heterogeneity in the treatment effect during the first few hours of work, with results seen below.

\begin{stlog}
. distcomp ylib , by(treated) alpha(0.05) p noplot
Comparing distribution of ylib when treated=0 vs. treated=1
 
Global test of equality of two CDFs:
    Simulated p-value = .14105487
    At a 10\% level: do not reject
    At a  5\% level: do not reject
    At a  1\% level: do not reject
{\smallskip}
. distcomp yfun , by(treated) a(0.05) p noplot
Comparing distribution of yfun when treated=0 vs. treated=1
 
Global test of equality of two CDFs:
    Simulated p-value = .03979024
    At a 10\% level: reject
    At a  5\% level: reject
    At a  1\% level: do not reject
 
With strong control of FWER at a 5\% level,
CDF equality is rejected at all points in the following
ranges of yfun:
{\smallskip}
     from         to 
        8         14 
\end{stlog}

For the library task, although the sample values look very different, the sample sizes are too small for the differences to be statistically significant at a $5\%$ FWER level (two-sided).
The FWER level would have to be $14.1\%$ (the GOF $p$-value) before rejecting equality in any range.
% $[56,58]$, near the upper end of the distribution.

For the fundraising task, even though the sample sizes are again small, the treatment effect is statistically significant at a $5\%$ level (two-sided).
The $p$-value is $0.040$ for CDF equality.
More specifically, \stcmd{distcomp} identifies the range of $\$8$ to $\$14$ as statistically significant at a $5\%$ FWER level.
(With $10\%$ FWER, the range $\$21$ to $\$26$ is also significant.)
This range is near the bottom of the distribution.
Opposite the library data entry task, where the gift wage treatment seemed to have the biggest effect on the upper end of the productivity distribution, the gift wage seems to have the biggest effect on the bottom end of the productivity distribution for door-to-door fundraising.

\begin{figure}[htbp]
\centering
\epsfig{file=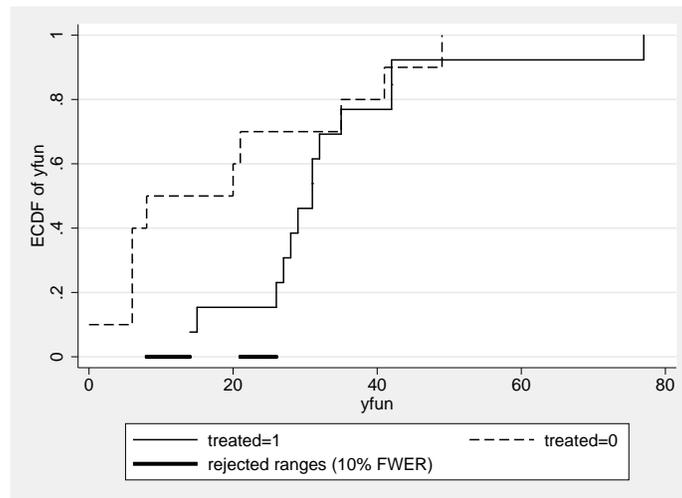}
\caption{\label{fig:yfun}Empirical CDFs from experiment (fundraising).}
\end{figure}

\begin{figure}[htbp]
\centering
\epsfig{file=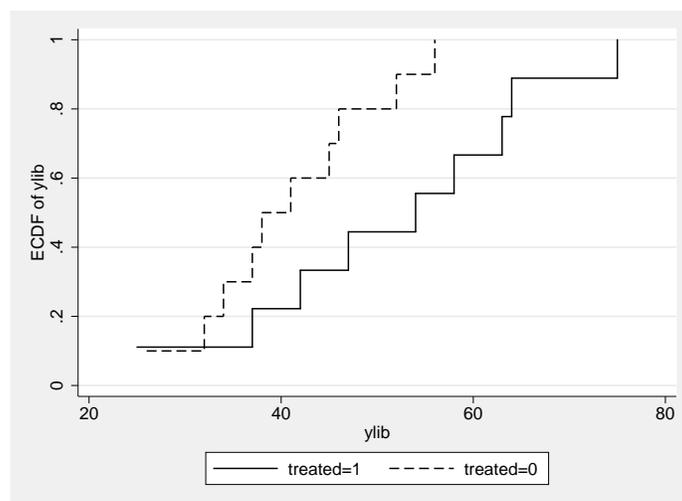}
\caption{\label{fig:ylib}Empirical CDFs from experiment (library data entry).}
\end{figure}

Figures \ref{fig:yfun} and \ref{fig:ylib} show the fundraising and library data entry empirical CDFs, respectively.
These are again the graphs generated automatically by \stcmd{distcomp}.
These graphs show the direction of the gift wage effect (higher productivity), but without \stcmd{distcomp} it is unclear where the differences are statistically significant.

\subsection{Example with regression discontinuity}
\label{sec:ex-RD}

The following regression discontinuity example uses data from \citet*{CattaneoEtAl2015}.
A longer version (with results from R code) appears in \citet[\S8.2]{GoldmanKaplan2018c}.
In brief, the research question is about the benefit of incumbency in U.S.\ Senate elections.
The regression discontinuity idea is essentially to consider elections where the incumbent won the prior election by a very small margin.
\Citet*{CattaneoEtAl2015} discuss a balance test-based bandwidth selection that suggests \texttt{h=0.75} percentage points is a small enough margin of victory that the outcome is (almost) as good as randomized.

In the following code and results, \texttt{demmv} is the Democratic margin of victory in the previous election for some Senate seat (in percentage points), which is negative if the Republican won.
Thus, the incumbent is a Democrat if \texttt{demmv} exceeds the threshold \texttt{R0=0}.
Also, \texttt{demvoteshfor2} is the Democratic vote share in the current election for the same Senate seat.
Below, the distribution of Democratic vote share is compared when the incumbent is a Democrat to when the incumbent is a Republican, restricting to cases where the incumbent's election was determined by a $0.75$ point or smaller margin of victory.

\begin{stlog}
. insheet using "https://sites.google.com/site/rdpackages/rdlocrand
> /r/rdlocrand_senate.csv",clear
(14 vars, 1,390 obs)
{\smallskip}
. scalar h = 0.75
{\smallskip}
. scalar R0 = 0
{\smallskip}
. gen D_incumbent = (demmv>=R0)
{\smallskip}
. distcomp demvoteshfor2 if demmv>=R0-h \& demmv<=R0+h , by(D_incumb
> ent) a(0.10) p
Comparing distribution of demvoteshfor2 when D_incumbent=0 vs. D_in
> cumbent=1
 
Global test of equality of two CDFs:
    Simulated p-value = .00721501
    At a 10\% level: reject
    At a  5\% level: reject
    At a  1\% level: reject
 
With strong control of FWER at a 10\% level,
CDF equality is rejected at all points in the following
ranges of demvoteshfor2:
{\smallskip}
     from         to 
 43.21114   47.73871 
 47.81345   51.70049 
 51.83689   56.64765 
\end{stlog}

The results show the incumbency effect to be statistically significant across most of the distribution.
With only two slim gaps, equality of the vote share distributions is rejected over the range from $43.2\%$ to $56.6\%$ of the vote.
Of course, beyond statistical significance, it is also important to estimate the magnitude of the incumbency effect, by the usual regression discontinuity estimator.

\begin{figure}[htbp]
\centering
\epsfig{file=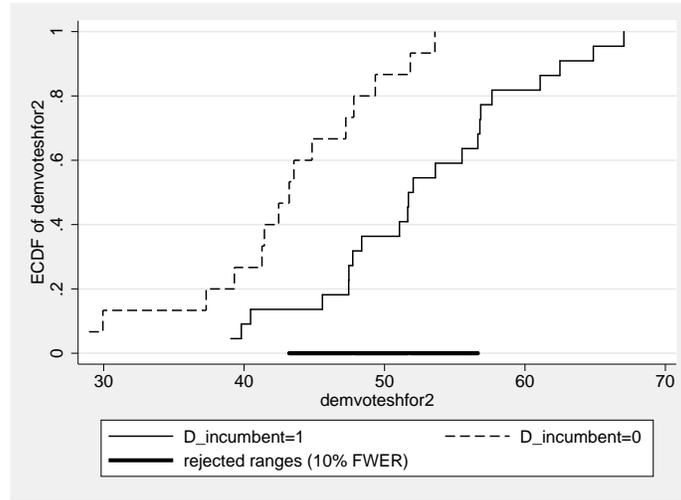}
\caption{\label{fig:RD}Empirical CDFs from regression discontinuity design.}
\end{figure}

Figure \ref{fig:RD} shows that in the sample, the vote share distribution for incumbents (who had a very small margin of victory) first-order stochastically dominates the distribution for challengers (i.e., non-incumbents).
It looks like with a larger sample size, equality could be rejected over an even larger range.
The graph also gives a sense of the magnitude of the incumbency effect.

\section{Methods and formulas}
\label{sec:methodology}

This section contains additional theoretical details from \citet{GoldmanKaplan2018c}.
It may provide a deeper understanding for some readers, although it may also be skipped without hindering successful application of \stcmd{distcomp} in practice.

Notationally, let $F_X(\cdot)$ be the population CDF for the first group, and let $F_Y(\cdot)$ be the population CDF for the second group.
The GOF null hypothesis is $H_0 \colon F_X(\cdot)=F_Y(\cdot)$, i.e., $F_X(r)=F_Y(r)$ for all real numbers $r$.
The \stcmd{distcomp} command also considers the individual hypotheses $H_{0r} \colon F_X(r)=F_Y(r)$.

For more formal discussion, the following notation and definitions are helpful.
These are adapted from the section on multiple testing by \citet[\S9.1]{LehmannRomano2005text}.
For the family of null hypotheses $H_{0r}$ indexed by real numbers $r$, let $\mathcal{T} \equiv \{r : H_{0r}\textrm{ is true}\}$, the set of values of $r$ for which hypothesis $H_{0r}$ is true.
The ``familywise error rate'' (FWER) is the probability of falsely rejecting at least one true hypothesis:
\begin{equation}
\FWER \equiv \Pr\left(\textrm{reject any $H_{0r}$ with $r \in \mathcal{T}$}\right) .
\end{equation}
``Weak control'' of FWER at level $\alpha$ requires $\FWER \le \alpha$ if each $H_{0r}$ is true, i.e., if the GOF null hypothesis $H_0 \colon F_X(\cdot)=F_Y(\cdot)$ is true, but it allows $\FWER>\alpha$ if some $H_{0r}$ are false.
In this setting, weak control of FWER is equivalent to size control for the corresponding GOF test that rejects $H_0 \colon F_X(\cdot)=F_Y(\cdot)$ when at least one $H_{0r}$ is rejected.
``Strong control'' of FWER requires $\FWER \le \alpha$ for any $\mathcal{T}$, i.e., for any two CDFs $F_X(\cdot)$ and $F_Y(\cdot)$.
Strong control implies weak control, but not vice-versa.

Strong control of FWER, even in small samples, is achieved by \stcmd{distcomp}.
The ``rejected ranges'' displayed by \stcmd{distcomp} are the values of $r$ for which $H_{0r}$ is rejected when strongly controlling the FWER at the specified level $\alpha$.

The two most important properties of \stcmd{distcomp} are its strong control of finite-sample FWER and its improved sensitivity to tail differences compared to \stcmd{ksmirnov}.
The procedure is now described mathematically, and the achievement of these two properties is then discussed further.

Steps for computation of \stcmd{distcomp} are given in Method 5 of \citet{GoldmanKaplan2018c}.
The idea is to compute a uniform confidence band (detailed below) for each unknown CDF, and then reject $H_{0r}$ for any $r$ where the bands do not overlap.
Notationally, let $B^{\tilde\alpha}_{k,n}$ denote the $\tilde\alpha$-quantile of the $\BetaDist(k,n+1-k)$ distribution, defining $B^{\tilde\alpha}_{0,n}=0$ and $B^{\tilde\alpha}_{n+1,n}=1$ for any $\tilde\alpha$, and denote sample sizes as $n_X$ and $n_Y$.
The uniform confidence bands for $F_X(\cdot)$ and $F_Y(\cdot)$ are respectively $[\sthat\ell_X(\cdot),\sthat{u}_X(\cdot)]$ and $[\sthat\ell_Y(\cdot),\sthat{u}_Y(\cdot)]$, where for some $\tilde\alpha$,
\begin{equation}
\sthat\ell_X(r) = B^{\tilde\alpha}_{n_X \sthat{F}_X(r), n_X} ,
\quad
\sthat{u}_X(r)  = B^{1-\tilde\alpha}_{n_X \sthat{F}_X(r)+1,n_X} ,
\end{equation}
and similarly replacing $X$ with $Y$.
Then, $H_{0r}$ is rejected when either $\sthat\ell_X(r)>\sthat{u}_Y(r)$ or $\sthat\ell_Y(r)>\sthat{u}_X(r)$.
The value of $\tilde\alpha$ is the largest value such that the probability of rejecting any $H_{0r}$ (i.e., the FWER) does not exceed $\alpha$ when $X_i \iid \UnifDist(0,1)$, $i=1,\ldots,n_X$, and $Y_i \iid \UnifDist(0,1)$, $i=1,\ldots,n_Y$, which is determined by simulation.

Although $\tilde\alpha$ is chosen to guarantee FWER control only when both $F_X(\cdot)$ and $F_Y(\cdot)$ are standard uniform CDFs, this extends to any $F_X(\cdot)=F_Y(\cdot)$.
The key insight is that at a given $r$, after determining $\tilde\alpha$ (and $n_X$ and $n_Y$), rejection of $H_{0r}$ depends only on $\sthat{F}_X(r)$ and $\sthat{F}_Y(r)$, which are step function that only increase at observed sample values.
That is, rejection of $H_{0r}$ only depends on the number of $X_i$ below $r$ and the number of $Y_i$ below $r$, which yield $\sthat{F}_X(r)$ and $\sthat{F}_Y(r)$ when divided by $n_X$ and $n_Y$, respectively.
Consequently, whether or not any $H_{0r}$ is rejected depends only on the relative order of observed $X_i$ and $Y_i$ values, not on the values themselves.
This implies that applying any monotonic transformation to the data does not affect the FWER.
When $F_X(\cdot)=F_Y(\cdot)=F(\cdot)$, we could sample $X_i$ and $Y_i$ from $F$ by first drawing standard uniform random variables and then applying $F^{-1}(\cdot)$.
But since $F^{-1}(\cdot)$ is a monotonic transformation, it will not affect FWER, so any $F(\cdot)$ will produce identical FWER as when $X_i$ and $Y_i$ are simply standard uniform themselves.

The above argument only concerns weak control of FWER; the extension to strong control is not obvious.
Indeed, one of the contributions of \citet{GoldmanKaplan2018c} is their Lemma 2 that proves weak control implies strong control for any procedure where rejection of $H_{0r}$ depends only on $\sthat{F}_X(r)$ and $\sthat{F}_Y(r)$, which is the case here.
The intuition is that if FWER is controlled when $F_X(r)=F_Y(r)$ for all $r$, then changing $F_X(\cdot)$ so that $F_X(r) \ne F_Y(r)$ at some $r$ does not somehow increase the probability of rejecting the remaining $H_{0r}$ where $F_X(r)=F_Y(r)$.

Computationally, the difficult part of the procedure is determining $\tilde\alpha$.
Since $\tilde\alpha$ depends only on $\alpha$, $n_X$, and $n_Y$, a large table of precise values was simulated ahead of time for the most common levels of $\alpha=0.01,0.05,0.10$, and included in \stcmd{distcomp}.
This enables nearly instantaneous computation of \stcmd{distcomp}.

For the second important property of improved tail sensitivity, it is insightful to look at the uniform confidence bands more closely.
Here, we look at a single sample of $X_i$ with CDF $F(\cdot)$.
A ``uniform confidence band'' for $F(\cdot)$ consists of an upper function $\sthat{u}(\cdot)$ and lower function $\sthat{\ell}(\cdot)$ that may depend on the data and satisfy $\Pr(\sthat{\ell}(\cdot) \le F(\cdot) \le \sthat{u}(\cdot)) \ge 1-\alpha$ for confidence level $(1-\alpha)\times100\%$, where $\sthat{\ell}(\cdot) \le F(\cdot) \le \sthat{u}(\cdot)$ means $\sthat{\ell}(r) \le F(r) \le \sthat{u}(r)$ for all $r$.
Such a band may be constructed by inverting the one-sample Kolmogorov--Smirnov test, but its pointwise coverage probability varies greatly with $r$.
That is, $\Pr(\sthat\ell(r) \le F(r) \le \sthat{u}(r))$ is much larger (closer to $100\%$) when $r$ is in the tails of the true distribution (i.e., when $F(r)$ is nearer zero or one) than when $r$ is in the middle (i.e., $F(r)$ nearer $0.5$).
In contrast, the pointwise coverage probability of the uniform confidence band used in \stcmd{distcomp} is (nearly) the same for all values of $r$.

The even pointwise coverage probability property of the uniform confidence bands used by \stcmd{distcomp} can be seen as follows.
Similar points are made by \citet[top p.\ 28]{BujaRolke2006}.
Let $X_{n:k}$ denote the $k$th order statistic in a sample of size $n$, i.e., the $k$th smallest value in the sample, so $X_{n:1} < \cdots < X_{n:k} < \cdots < X_{n:n}$.
From \citet{Wilks1962}, $F(X_{n:k}) \sim \BetaDist(k, n+1-k)$ if $F(\cdot)$ is continuous.
This follows from an application of the ``probability integral transform'': $F(X_i) \iid \UnifDist(0,1)$, and $F(X_{n:k})$ follows the same distribution as the $k$th order statistic from a sample of $n$ standard uniform random variables, which is $\BetaDist(k, n+1-k)$.
Thus, since $B^{\tilde\alpha}_{k,n}$ is defined as the $\tilde\alpha$-quantile of that same distribution, $\Pr( B^{\tilde\alpha}_{k,n} \le F(X_{n:k}) \le B^{1-\tilde\alpha}_{k,n} ) = 1-2\tilde\alpha$ exactly, for any $k$ and $n$, irrespective of $F(\cdot)$.
In the earlier expressions, $[\sthat\ell(\cdot),\sthat{u}(\cdot)]$ is essentially taking pointwise intervals $[\sthat\ell(X_{n:k}),\sthat{u}(X_{n:k})]=[B^{\tilde\alpha}_{k,n},B^{1-\tilde\alpha}_{k,n}]$ and connecting them with a stair-step interpolation.
This implies pointwise coverage probability of $1-2\tilde\alpha$ at every order statistic (and only somewhat larger at other points).

The $1-2\tilde\alpha$ probability at each order statistic contrasts the Kolmogorov--Smirnov pointwise coverage probability that is much higher in the tails.
This difference translates directly to the ability to detect deviations across different values: the Kolmogorov--Smirnov sensitivity/power is concentrated in the center of the distribution, whereas \stcmd{distcomp} spreads its power evenly across the whole distribution.
Put differently, Kolmogorov--Smirnov implicitly uses a much larger pointwise statistical significance level for testing $H_{0r}$ near the center of the distribution and much smaller significance level in the tails, whereas \stcmd{distcomp} uses approximately the same level of statistical significance for all $H_{0r}$.

\section{Conclusion}
\label{sec:conc}

In addition to a two-sample goodness-of-fit test that improves upon the Kolmogorov--Smirnov test, the \stcmd{distcomp} command provides a detailed, point-by-point assessment of statistically significant differences between two distributions.
This is much more informative than existing goodness-of-fit tests (like the \stcmd{ksmirnov} command) or $t$-tests for mean equality (like \stcmd{ttest}) while still controlling the false positive rate, with strong control of the familywise error rate.
Potential applications abound, such as descriptions of how a variable's distribution changes over time or differs between groups (geographic, socioeconomic, etc.), regression discontinuity designs, and perhaps especially in program evaluation.

\section{Acknowledgements}
\label{sec:thx}

There would be no Stata command without the work of Matt Goldman (Facebook Research) on \citet{GoldmanKaplan2018c}.
Thanks to \citet{GneezyList2006} and \citet*{CattaneoEtAl2015} for making their data public.
Thanks to Colin Cameron for first encouraging me (in 2013) to write Stata commands.

\bibliographystyle{sj}
% \bibliography{_stata}
\ifnum 8=1 \def\bibname{Reference}
\else \def\bibname{References} \fi

\begin{aboutauthor}
David M.\ Kaplan is an assistant professor in the Department of Economics at the University of Missouri.
His primary research interest is econometric methodology.
In particular, he enjoys creating and advancing methods for understanding changes and treatment effects on entire distributions (instead of just averages).
\end{aboutauthor}

\clearpage
\end{document}